%% file: main.tex
\documentclass[10pt,conference]{IEEEtran}

\usepackage{cite}
\usepackage{amsmath,amssymb,amsthm, amsfonts}   
\usepackage{mathtools}
\usepackage{graphicx}
\usepackage{textcomp}
\usepackage{xcolor}
\usepackage[caption=false,font=footnotesize]{subfig}
\usepackage{stfloats}
\usepackage{epstopdf}
\usepackage{acronym}
\usepackage{mathrsfs}
\usepackage{float}
\usepackage{placeins}
\usepackage{flushend}
\usepackage{algorithm}
\usepackage[algo2e]{algorithm2e}
\usepackage{algpseudocode}
\usepackage{booktabs}
\usepackage{multicol}
\usepackage{svg}
\usepackage{subfig}
\usepackage{verbatim}
\usepackage{lipsum}
\usepackage[bookmarks=false]{hyperref}

\newtheorem{lemma}{Lemma}
\newtheorem{proposition}{Proposition}
\newtheorem{corollary}{Corollary}

\def\BibTeX{{\rm B\kern-.05em{\sc i\kern-.025em b}\kern-.08em
		T\kern-.1667em\lower.7ex\hbox{E}\kern-.125emX}}
        
\input{acronyms}
\input{variables}

\DeclareMathAlphabet{\pazocal}{OMS}{zplm}{m}{n}
\newcommand{\unif}{\pazocal{U}}

\IEEEoverridecommandlockouts

\begin{document}
	
    \title{A Smooshed BMOCZ Zero Constellation for CFO Estimation Without Channel Coding}
	
    \author{Anthony Joseph Perre, Parker Huggins, and Alphan \c{S}ahin\\
    Department of Electrical Engineering, University of South Carolina, Columbia, SC, USA\\
    Email: \{aperre, parkerkh\}@email.sc.edu, asahin@mailbox.sc.edu
 	\thanks{This work has been supported by the \ac{NSF} through the award CNS-2438837.}    
}
    
    \maketitle
    
    \begin{abstract}	
    In this study, we propose a new \ac{BMOCZ} zero constellation, which we call \ac{SBMOCZ}, to address \ac{CFO}-induced zero rotation without depending on channel coding. In our approach, we modify the phase mapping of Huffman \ac{BMOCZ} by shrinking the angle between adjacent zeros, except for the first and last, to introduce a gap in the zero constellation. By discerning the gap location in the received polynomial, the receiver can estimate and correct the phase rotation. We demonstrate the error rate performance of \ac{SBMOCZ} relative to Huffman \ac{BMOCZ}, showing that \ac{SBMOCZ} addresses a \ac{CFO}-induced rotation at the cost of a modest performance reduction compared to Huffman \ac{BMOCZ} in the absence of a \ac{CFO}. Finally, we compare \ac{SBMOCZ} to Huffman \ac{BMOCZ} using a \ac{CPC}, showing a $\mathbf{4}$~dB \ac{BER} improvement in a fading channel, while demonstrating comparable performance across other simulations.

    \end{abstract}

    \acresetall
    
\section{Introduction} \label{sec:intro}
Non-coherent communication has emerged as a promising means to simplify receiver design and lower power consumption for ultra-massive connectivity in wireless networks\cite{nawaz2021backscatter}. In particular, non-coherent communication offers unique advantages in wireless networks by eliminating the need for explicit \ac{CSI}, which makes it suitable for low data rate applications\cite{witrisal2009noncoherent}. The lack of overhead in non-coherent communication makes it especially beneficial in dynamic environments where \ac{CSI} estimation may be impractical\cite{xu2019tradeoffs}. Since non-coherent communication suffers from degraded performance compared to coherent communication, improving its reliability and spectral efficiency remains an active area of research.

In \cite{walk2019principles}, the authors propose a novel non-coherent modulation scheme called \ac{BMOCZ}, where the information bits are mapped to the zeros of the baseband signal's $z$-transform. The information zeros are constrained to lie on one of two concentric circles in the complex plane, forming a \textit{zero constellation} which the authors term \textit{Huffman \ac{BMOCZ}}. Previous studies have explored the optimal radius for \ac{BMOCZ}, and examined its integration into \ac{OFDM}\cite{huggins2024optimal}. The primary advantage of \ac{BMOCZ} is that the information zeros are preserved at the receiver regardless of the channel realization, thereby making it non-coherent. This makes \ac{BMOCZ} ideal for applications requiring ultra-reliable, low-latency communication, such as intermittent short-packet transmissions in \ac{IoT} networks\cite{walk2020practical}. One notable use is over-the-air computation, a technique that struggles with channel-induced distortions and requires \ac{CSI} estimation, which can be impractical in dynamic environments \cite{asahin2024majority}. Moreover, the desirable auto-correlation properties of \ac{BMOCZ} make it an ideal choice for integrated sensing and communication\cite{dehkordi2023integrated}. In general, \ac{BMOCZ} offers specific advantages as a non-coherent modulation scheme across a wide range of potential applications.

When utilizing \ac{BMOCZ} for wireless communication, there are two notable impairments that can degrade the error rate performance: a \ac{TO} and \ac{CFO} \cite{walk2020practical}. A \ac{TO} occurs whenever the start time of the transmitted signal misaligns with its actual arrival, leading to \ac{ISI}. In \cite{walk2020practical}, the authors exploit the auto-correlation properties of Huffman \ac{BMOCZ} to estimate and correct the \ac{TO}. The \ac{CFO} can also significantly degrade the error rate performance. To address the \ac{CFO} in \ac{BMOCZ}-based communication, the authors in \cite{walk2020practical} introduce an \ac{ACPC} combined with an oversampled \ac{dizet} decoder to estimate and correct the \ac{CFO}. However, the \ac{ACPC} limits the code structure to \textit{\acp{CPC}}. Therefore, to improve flexibility, alternative \ac{CFO} correction methods for \ac{BMOCZ} must be explored.

In this study, we propose a new smooshed zero constellation called \ac{SBMOCZ}, where the angular separation between adjacent zeros, excluding the first and last, is decreased. This creates a distinct \textit{gap} in the zero constellation, which will rotate under a \ac{CFO}. By identifying the gap location in the received polynomial, we can correct the \ac{CFO} \textit{without channel coding}. We highlight that our algorithm for estimating a \ac{CFO}-induced rotation can be implemented via a single $N$-point \ac{DFT}, making the implementation relatively efficient. The simulation results for \ac{BER} and \ac{BLER} in \ac{AWGN} and fading channels demonstrate that \ac{SBMOCZ} functions under a \ac{CFO} without any channel coding, while incurring a modest performance loss compared to Huffman \ac{BMOCZ} without a \ac{CFO}. Furthermore, comparisons of coded \ac{SBMOCZ} and Huffman \ac{BMOCZ} with an \ac{ACPC} show that \ac{SBMOCZ} achieves a $4$~dB \ac{BER} gain in the fading channel, while maintaining similar performance across other scenarios.

\emph{Notation}: The sets of real and complex numbers are denoted by $\realnumbers$ and $\complexnumbers$, respectively. We define $\integers^N_2$ as the set of all binary vectors of length $N$. The imaginary unit is given by $\j=\sqrt{-1}$, and Euler's number is represented with $\e$. The complex conjugate of $z=a+\j b$ is expressed as $z^\ast=a-\j b$. We denote the Euclidean norm of a vector $\vector\in\complexnumbers^{N\times1}$ as $||\vector||_2=\sqrt{\vector\herm\vector}$. The circularly symmetric complex normal distribution with mean zero and variance $\sigma^2$ is expressed as $\complexnormal(0,\sigma^2)$. The uniform distribution on the interval $\left[a, b\right)$ is denoted by $\unif_{[a, b)}$. We define $[N]=\{0,1,\dots,N-1\}$ to represent the set of the first $N$ non-negative integers. Finally, we express the floor of a real number $a$ as $\lfloor a \rfloor$.


\section{System model} \label{sec:model}

	\subsection{BMOCZ Fundamentals}
	For a \ac{BMOCZ} constellation, the $k$th bit in a binary message $\messageSeq=[m_0,m_1,\hdots,m_{\numZeros-1}]\in \integers^K_2$ is mapped to the $k$th zero of a polynomial according to
	\begin{equation} \label{eq:zero_mapping}
		\zero_k = 
		\begin{cases}
			\radius_k \, \e^{\j\phi_k}, & m_\zeroIndex=1\\
			\radius_k^{-1} \, \e^{\j\phi_k}, & m_\zeroIndex=0
		\end{cases}\:, \ \ k\in[\numZeros] \;,
	\end{equation} with $\radius_k>1$, and $\phi_k \in \left[0, 2\pi \right)$. By the fundamental theorem of algebra, the zeros $\zeroVector=[\zero_0,\zero_{1},\dots,\zero_{\numZeros-1}] \in \complexnumbers^\numZeros$ uniquely define the $\numZeros$th degree polynomial
	\begin{equation} \label{eq:Xpolynomial}
		X(z)=\sum_{k=0}^{\numZeros}x_\zeroIndex z^\zeroIndex=x_\numZeros \prod_{\zeroIndex=0}^{\numZeros-1}(z-\zero_\zeroIndex) \;,
	\end{equation} where $x_\numZeros\neq0$. The polynomial coefficients, given by  
	$\polySeqTX = [x_0, x_1, \dots, x_K] \in \complexnumbers^{\numZeros+1}$, are normalized such that $||\polySeqTX||_2^2=\numZeros+1$. Let $W(z) = \sum_{n=0}^{\zerosAndTaps-1} w_n z^n$ and $H(z) = \sum_{l=0}^{\numTaps-1} h_l z^l$ indicate the $z$-domain representations for the noise sequence  
	$\noiseSeq = [w_0, w_1, \dots, w_{\zerosAndTaps-1}] \in \mathbb{C}^{\zerosAndTaps}$ and the $\numTaps$-tap channel impulse response  
	$\filter=[h_0, h_1, \dots, h_{\numTaps-1}]\in\mathbb{C}^{\numTaps}$, respectively. Assuming transmission through an LTI channel and applying the convolution theorem for $\zerosAndTaps=\numZeros+\numTaps$, the received sequence can be expressed in the $z$-domain as  
	\begin{equation} \label{eq:z_domain_mult}
		Y(z)=\sum_{n=0}^{\zerosAndTaps-1} y_n z^n=X(z)H(z)+W(z) \;,
	\end{equation} where $\polySeqRX = [y_0, y_1, \dots, y_{\zerosAndTaps-1}]\in\complexnumbers^{\zerosAndTaps}$ is the vector containing the coefficients of $Y(z)$. Although the presence of $H(z)$ introduces $L-1$ additional zeros to $Y(z)$, it does not alter the zeros corresponding to the message in $X(z)$. This property allows \ac{BMOCZ} to operate without \ac{CSI}, thereby making it \textit{non-coherent}. In this study, we assume a flat-fading channel where $L=1$, which means $\zerosAndTaps=K+1$. This can be achieved, for example, through a time-frequency mapping of \ac{BMOCZ} polynomial coefficients in \ac{OFDM}, which ensures $X(z)$ and $Y(z)$ have the same number of zeros \cite{huggins2024optimal}.

	In \cite{walk2019principles}, the authors introduce \textit{Huffman \ac{BMOCZ}}, where the information zeros are positioned on one of two concentric circles in the complex plane. This results in the zero mapping rule provided in \eqref{eq:zero_mapping}, with $R_k = r_{\text{hb}}$ and $\phi_k \triangleq 2\pi k/\numZeros$. In this scheme, $X(z)$ is called a \textit{Huffman polynomial}, since the coefficients form a Huffman sequence for any combination of information zeros \cite{ackroyd1970design, walk2017short}. Huffman polynomials are well-conditioned, meaning that small variations in the coefficients lead to slight changes in the zeros, making them ideal for communication systems. In \cite{walk2019principles}, the authors derive a simple decoding rule for \ac{BMOCZ} called \ac{dizet}, which evaluates $Y(z)$ at each conjugate-reciprocal zero pair $\zeroCodebook_k\triangleq \{\zero_k,1/\zero_k^\ast\}$. Using this method, the $k$th detected bit is given by
	\begin{equation} \label{eq:dizet_decoder}
		\estMessageBit_k = 
		\begin{cases}
			1, & \left|Y(\radius_k \, \e^{\j\phi_k})\right| < \radius_k^{\zerosAndTaps-1}\left|Y(\radius_k^{-1} \, \e^{\j\phi_k})\right|\\
			0, & \text{otherwise}
		\end{cases} \;.
	\end{equation} To control the minimum pairwise separation between zeros, the radius is defined as
	\begin{equation} \label{eq:dizet_radius}
		r_{\text{hb}} \triangleq \sqrt{1+2\lambda\sin{\left(\pi/\numZeros\right)}}\:.
	\end{equation} According to \cite{walk2019principles}, radial zero separation has a stronger influence on \ac{BER} than angular separation. Therefore, the parameter $\lambda\in\left(0,1\right]$ is introduced as a trade-off factor to balance radial versus angular zero separation.

	\subsection{CFO Impairment}
	A \ac{CFO} occurs due to frequency mismatches between the transmitter and receiver oscillators. The presence of a \ac{CFO} in Huffman \ac{BMOCZ} is a significant concern because it degrades the system performance. Let $\psi \in \left[0, 2\pi\right)$ denote the amount of phase rotation caused by the \ac{CFO}.\footnote{Although this range of \ac{CFO}-induced rotation is unrealistic in practice, and could disrupt the subcarrier orthogonality in \ac{OFDM}, we consider it here for consistency with \cite{walk2020practical}.} Due to this impairment, $Y(z)$ experiences a transformation in the $z$-domain
	\begin{equation} \label{eq:polynomial_cfo}
		\tilde{Y}(z)=\sum_{n=0}^{\zerosAndTaps-1}y_n \e^{\j\psi n} z^n= Y(z\e^{\j\psi}) \;.
	\end{equation} Hence, the zeros of $Y(z)$ rotate clockwise by the angle $\psi$. For a Huffman \ac{BMOCZ} constellation with small $\numZeros$, this issue is less significant. However, for moderate or large values of $\numZeros$, even a small rotation causes decoding failure. Huffman \ac{BMOCZ}, in the absence of channel coding, can only correct a fractional \ac{CFO} due to the constant phase separation $\Delta{\phi} = 2\pi/K$ between adjacent zeros and the uniform radii. Therefore, an angular rotation $\psi$ is only unique modulo $\Delta{\phi}$. Decomposing $\psi$ into a fractional component $\psi_0$ and an integer multiple of $\Delta{\phi}$, we obtain
	\begin{equation} \label{eq:fractional_cfo_psi}
	\psi = \psi_0 + m\Delta{\phi} \; \; \text{where} \; \;\psi_0\in\left[0, \Delta{\phi}\right) \;,
	\end{equation} for any $m\in[K]$. Therefore, only the fractional component $\psi_0$, relative to the angular separation $m\Delta{\phi}$, is detectable with uncoded Huffman \ac{BMOCZ}. To address this issue, the authors in \cite{walk2020practical} propose an \ac{ACPC} combined with an oversampled \ac{dizet} decoder to estimate and correct the total zero rotation. Within this framework, the oversampled \ac{dizet} decoder estimates the fractional component $\psi_0$, while the \ac{ACPC} identifies the cyclic shift $m\Delta{\phi}$.
	
	We highlight several limitations associated with the \ac{ACPC}. First, we note that the complexity of the code construction for an \ac{ACPC} increases when the code length is not a Mersenne prime, i.e., a prime number of the form $2^K-1$ for prime $K$. Furthermore, the approach requires using \textit{cyclic codes}, which confines the range of available coding structures. For instance, integrating \ac{BMOCZ} with \ac{LDPC} or polar codes is not practical within the \ac{ACPC} framework. Finally, the approach introduces extra computational overhead in the decoding of the chosen cyclic code, which is required to correct the zero rotation. For a detailed overview of \ac{ACPC}, we direct the reader to \cite{walk2020practical}. In this work, to retain flexibility, we propose a new \ac{BMOCZ} zero constellation to address \ac{CFO}-induced rotation \textit{without} any channel coding.
	
\section{Methodology} \label{sec:methods}
The following subsections introduce a framework to estimate and correct \ac{CFO}-induced rotation by using a smooshed \ac{BMOCZ} zero constellation.
	
    \subsection{Smooshed Zero Constellation} \label{subsec:modified_zero_constellation}
   	We introduce a smooshed \ac{BMOCZ} zero constellation where the phase difference between adjacent zeros, excluding the first and last, is given by $\Delta{\phi}=(2\pi-\smooshFactor)/K$. The parameter $\smooshFactor\in\left[0, 2\pi\right)$, referred to as the \textit{smooshing factor}, compresses the phase separation, which creates a larger \textit{gap} between $\alpha_0$ and $\alpha_{K-1}$. The evaluation of $|X(z)|$ along the unit circle is maximized closest to the center of the gap, since the density of zeros is lower in that region. As the constellation rotates, the gap shifts, which causes the maximum to move accordingly. Therefore, by evaluating the rotated polynomial $\tilde{Y}(z)$ at $z=\e^{-\j\theta}$ for $\theta\in[0, 2\pi)$, we can estimate $\psi$ from the $\theta$ that maximizes $|\tilde{Y}(\e^{-\j\theta})|$. A detailed explanation for choosing the unit circle, along with a discussion of the algorithm for \ac{CFO} estimation and correction, is provided in Section \ref{subsec:CFO_estimation}.
    
    We choose to center the gap on the positive real axis, which yields a new phase mapping
    \begin{equation} \label{eq:modified_phase_mapping}
    	\phi_k \triangleq \frac{(2\pi - \smooshFactor)\zeroIndex}{\numZeros} + \frac{2\pi+\smooshFactor\left(\numZeros-1\right)}{2\numZeros} \;.
    \end{equation} The selection of $\smooshFactor$ affects the reliability of \ac{CFO} correction and the displacement of zeros under noise. In particular, a small $\smooshFactor$ results in poor \ac{CFO} correction capabilities, while an excessively large $\smooshFactor$ leads to significant zero perturbation under noise. Therefore, careful selection of the smooshing factor is required to obtain good error rate performance. The choice of $\smooshFactor$ also affects the minimum pairwise separation of the zeros, and hence the selection of radius. We modify \eqref{eq:dizet_radius} and obtain
    \begin{equation} \label{eq:modified_radii}
     	r_{\text{sb}} \triangleq \sqrt{1+2\lambda\sin{\left(\frac{2\pi - \smooshFactor}{2\numZeros}\right)}} \;.
    \end{equation} A derivation for \eqref{eq:modified_radii} is given in Appendix \ref{app:sbmoczradii}.
    
\begin{figure}[t]
    	\centering
    	\subfloat[Huffman \ac{BMOCZ}.]{\includegraphics[width=2.477in]{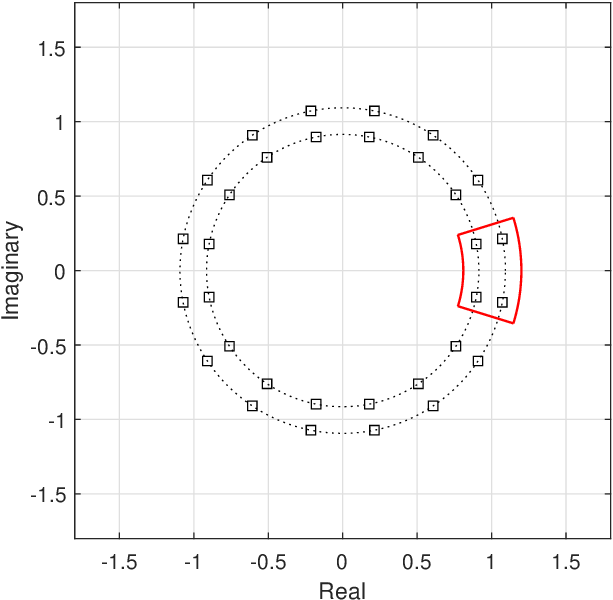}} \\
    	\subfloat[\ac{SBMOCZ} with $\smooshFactor=1/2$.]{\includegraphics[width=2.477in]{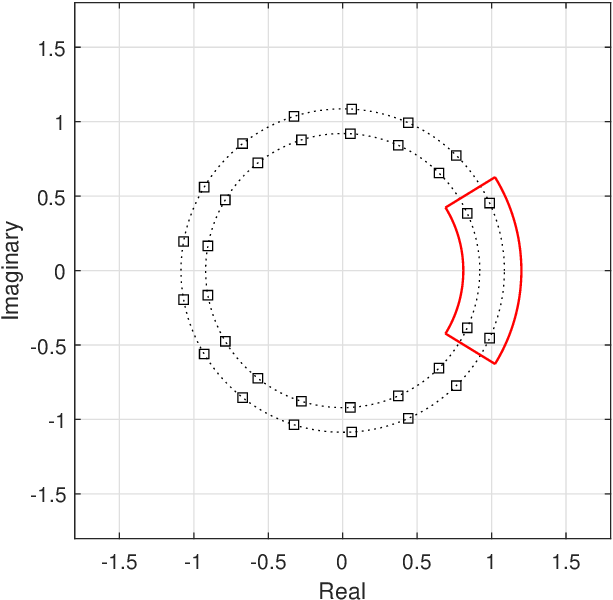}}
    	\caption{Comparison of Huffman \ac{BMOCZ} and \ac{SBMOCZ} for $\numZeros=16$. The square markers indicate possible zero locations for the transmitted polynomial $X(z)$.}
    	\label{fig:constellation_comparison}
    \end{figure}

    When $\smooshFactor=0$, observe that \eqref{eq:modified_phase_mapping} and \eqref{eq:modified_radii} reduce to Huffman \ac{BMOCZ} with a $\pi/K$ counterclockwise rotation of the zeros. In this study, for simplicity, we treat \ac{SBMOCZ} with $\smooshFactor=0$ and Huffman \ac{BMOCZ} interchangeably. Fig. \ref{fig:constellation_comparison} illustrates the zero constellations for Huffman \ac{BMOCZ} and \ac{SBMOCZ} with $K=16$, where $\smooshFactor=1/2$ for \ac{SBMOCZ}.\footnote{While $\smooshFactor$ is typically much smaller in practice, we increase it here to emphasize the zero smooshing effect.} The radii are computed by setting $\lambda=1/2$ and applying \eqref{eq:dizet_radius} and \eqref{eq:modified_radii} to derive $\radius_k$ for Huffman \ac{BMOCZ} and \ac{SBMOCZ}, respectively.
    
    \begin{figure*}[t]
    	\centering
    	\subfloat[Huffman \ac{BMOCZ}.]{
    		\includegraphics[width=0.3\textwidth]{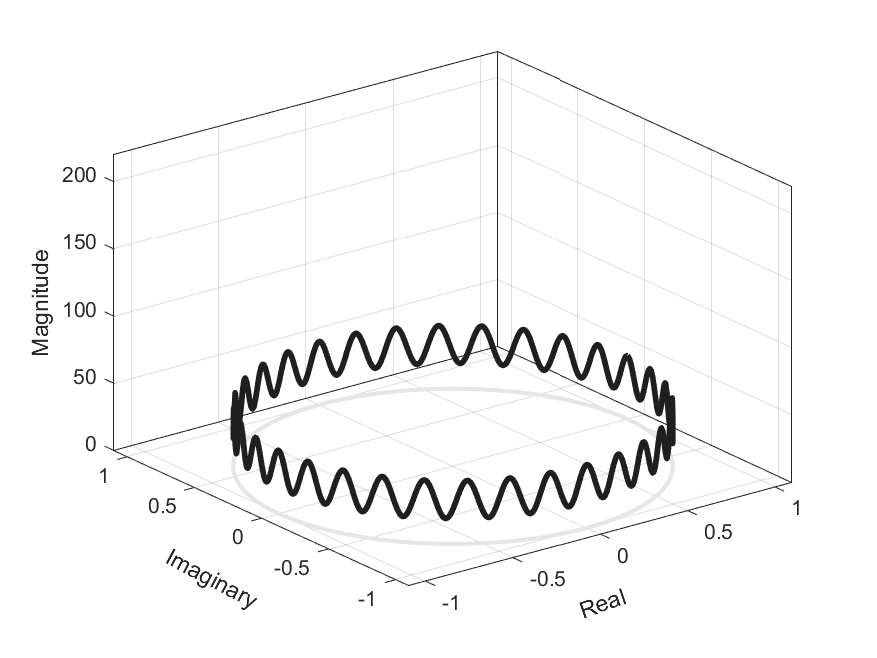}
    		\label{subfig:huffman}
    	}
    	\hfill
    	\subfloat[\ac{SBMOCZ} with $\smooshFactor=3/100$.]{
    		\includegraphics[width=0.3\textwidth]{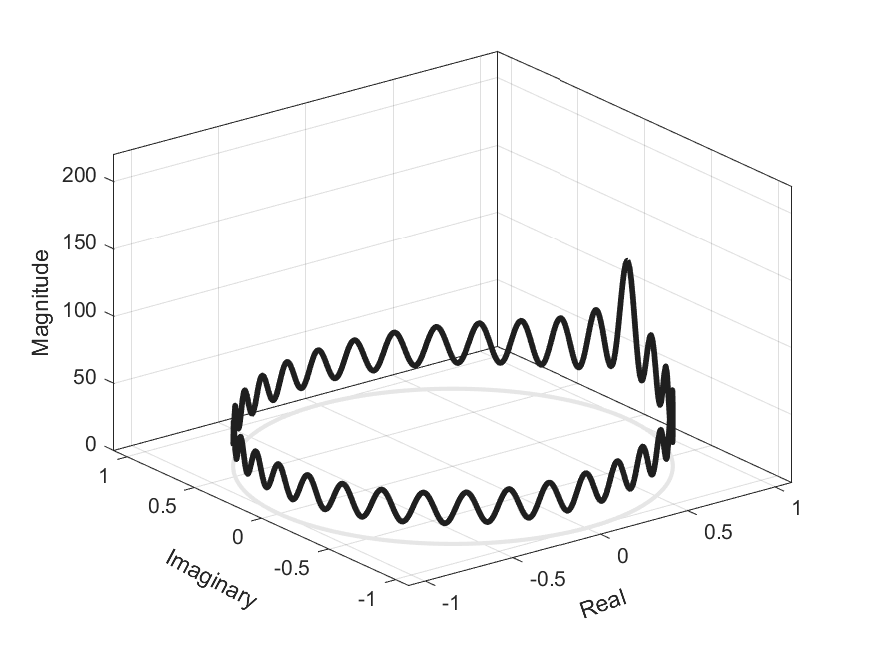}
    		\label{subfig:sbmocz_3e-2}
    	}
    	\hfill
    	\subfloat[\ac{SBMOCZ} with $\smooshFactor=1/20$.]{
    		\includegraphics[width=0.3\textwidth]{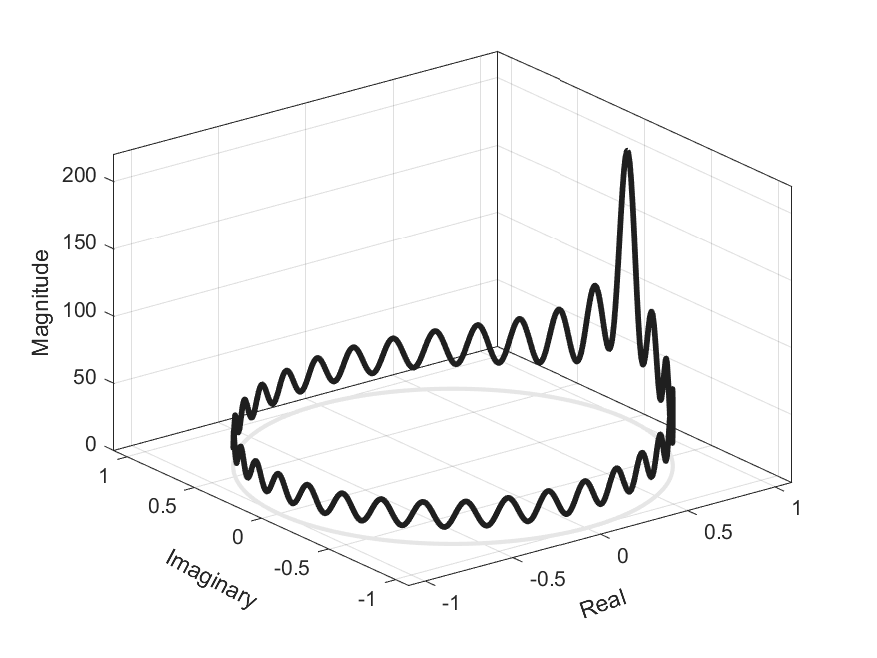}
    		\label{subfig:sbmocz_5e-2}
    	}
    	\caption{$|X(z)|^2$ evaluated on the unit circle for $K=32$.}
    	\label{fig:eta_peak_comparison}
    \end{figure*}
     
     \subsection{CFO Correction with \ac{SBMOCZ}}\label{subsec:CFO_estimation}
     In this subsection, we introduce a method to correct a \ac{CFO} impairment by evaluating the rotated polynomial $\tilde{Y}(z)$ at various points along the unit circle. To justify evaluating on the unit circle, we highlight some useful properties. We begin with the following proposition:
     
     \begin{proposition} \label{lemma:dtftauto}
     	Let $X(z)$ be a polynomial with the coefficient vector $\polySeqTX=[x_0, x_1, \dots, x_K] \in \complexnumbers^{\numZeros+1}$. Then, the squared magnitude of $X(\e^{-\j\theta})$ for $\theta \in \left[0, 2\pi\right)$ corresponds to the \ac{DTFT} of the auto-correlation sequence  $\bold{a} =[a_{-\numZeros},a_{-\numZeros+1}, \dots,a_{\numZeros}]\in\complexnumbers^{2\numZeros+1}$ for $\polySeqTX$. In particular, we have
   		\begin{align}
   			\left|X(\e^{-\j\theta})\right|^2 = \sum_{\aacfindex=-\numZeros}^{\numZeros} a_\aacfindex \e^{-\j\theta \aacfindex} \; .
   		\end{align}
     \end{proposition}
     
     This result follows from the fact that evaluating a polynomial on the unit circle yields its \ac{DTFT}, and that the squared magnitude corresponds to the power spectral density, which is the \ac{DTFT} of the auto-correlation sequence.
     
     \begin{corollary} \label{corr:dtftauto}
     	The auto-correlation sequence $\bold{a}\in\complexnumbers^{2\numZeros+1}$ is the same for any transmitted \ac{BMOCZ} coefficient sequence $\polySeqTX\in\complexnumbers^{\numZeros+1}$ \cite{walk2019principles}. Consequently, for any \ac{SBMOCZ} polynomial $X(z)$, the evaluation $|X(\e^{-\j\theta})|^2$ yields the same result.
     \end{corollary}
     
     The important takeaway from Proposition~\ref{lemma:dtftauto} and Corollary~\ref{corr:dtftauto} is that the smooshed zero constellation gap will induce a peak on the unit circle at the \textit{same} location for any transmitted \ac{SBMOCZ} coefficient sequence $\polySeqTX$. Additionally, evaluation on the unit circle will be \textit{symmetric} about the peak, as highlighted by the following lemma:
     
     \begin{lemma} \label{lemma:realval}
     	Let $X(z)$ be an \ac{SBMOCZ} polynomial. When evaluating $|X(\e^{-\j\theta})|^2$ with $\theta\in [0, 2\pi)$, we have
     	\begin{align}
     		|X(\e^{-\j\theta})|^2 = |X(\e^{\j\theta})|^2 \; .
     	\end{align}
     \end{lemma}
     The proof is given in Appendix~\ref{app:lemma}. For \ac{SBMOCZ}, the gap is centered on the positive real axis, resulting in a peak on the unit circle at $z=1$. To illustrate this behavior, we now present Fig. \ref{fig:eta_peak_comparison}, which plots $|X(z)|^2$ along the unit circle for different $\smooshFactor$ with $K=32$. As the gap increases in size, the peak becomes more pronounced, which makes the \ac{CFO} estimate more robust against noise. However, this comes at the price of larger zero perturbation under noise. Additionally, Fig. \ref{fig:eta_peak_comparison} shows that Huffman \ac{BMOCZ} is incompatible with our approach, since there is not a unique maximum on $\left[0, 2\pi\right)$. Instead, evaluation on the unit circle exhibits sinusoidal behavior for Huffman \ac{BMOCZ}. The key observation is that by identifying the peak location for $|\tilde{Y}(z)|^2$ on the unit circle, we can obtain an estimate for the angular rotation $\psi$. 
    
     Without loss of generality, we will now consider $|\tilde{Y}(z)|$, which peaks at the same location as its square. Under rotation, the maximum of $|X(\e^{\j(\theta+\psi)})|$ in \ac{SBMOCZ} occurs whenever $\theta=-\psi$, so we can estimate the angular rotation from $\tilde{Y}(z)$ via
     \begin{equation} \label{eq:cfo_estimation_continous}
    	 \hat{\psi} = \arg\! \! \! \, \max_{\theta \in \left[0, 2\pi\right)} \left|\tilde{Y}(\e^{-\j\theta})\right| \;.
     \end{equation} To discretize this process, we approximate $\hat{\psi}$ as
     \begin{equation} \label{eq:cfo_estimation_discrete}
    	 \hat{\psi}\approx \frac{2\pi}{N}\arg \max_{n\in[N]} \left|\tilde{Y}(\e^{-\j2\pi n/N})\right| \;,
     \end{equation} where $N$ determines the resolution of the search space. We highlight that $\tilde{Y}(\e^{-\j2\pi n/N})$ can be evaluated using a single $N$-point \ac{DFT} of the coefficients, which results in a modest time complexity of $\complexity(N\log N)$ for the proposed algorithm. To compensate for the rotation, we perform the correction $\hat{Y}(z) = \tilde{Y}(z\e^{-\j\hat{\psi}})$, where $\hat{Y}(z)$ is an estimate of the received polynomial without a \ac{CFO} impairment. The \ac{dizet} decoder is then applied to $\hat{Y}(z)$ to recover the binary message $\messageSeq\in\integers^K_2$.
    
    \begin{figure}[t!]
    	\centering
    	\subfloat[Bit error rate curves. \label{subfig:uncoded_performance_ber}]{\includegraphics[width=3in]{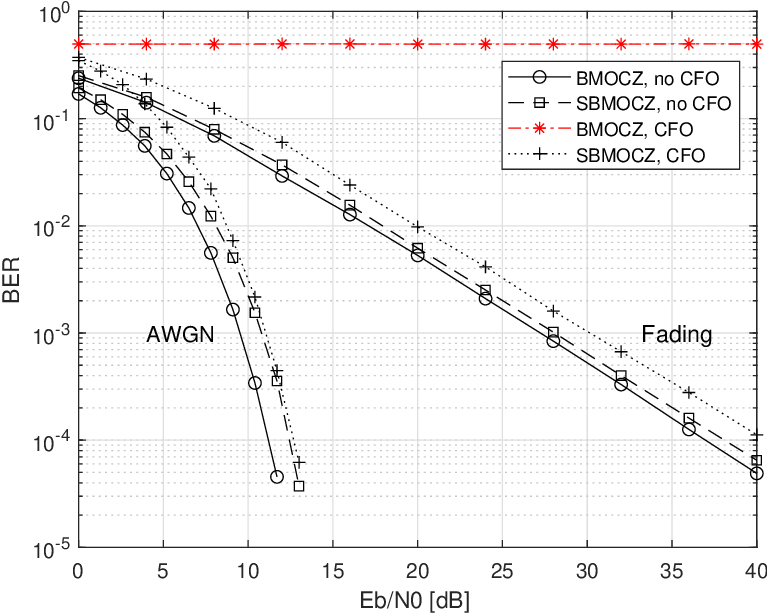}} \\
    	\subfloat[Block error rate curves. \label{subfig:uncoded_performance_bler}]{\includegraphics[width=3in]{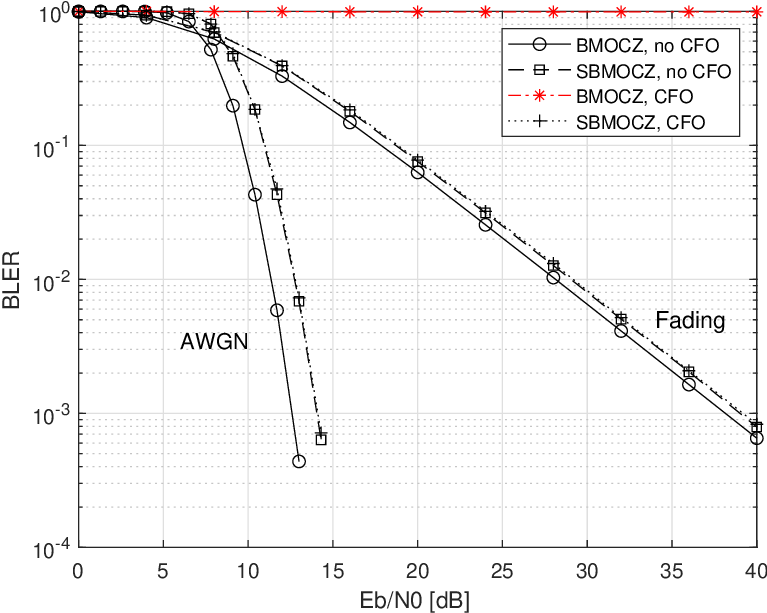}}
    	\caption{Comparison of uncoded schemes for $\numZeros=128$. The \ac{SBMOCZ} scheme is configured with $r_{\text{sb}}=1.0122$ and $\smooshFactor=0.0117$, while Huffman BMOCZ uses $r_{\text{hb}}=1.0122$.}
    	\label{fig:uncoded_performance}
    \end{figure}

    \begin{figure}[t!]
    	\centering
    	\subfloat[Bit error rate curves. \label{subfig:coded_performance_ber}]{\includegraphics[width=3in]{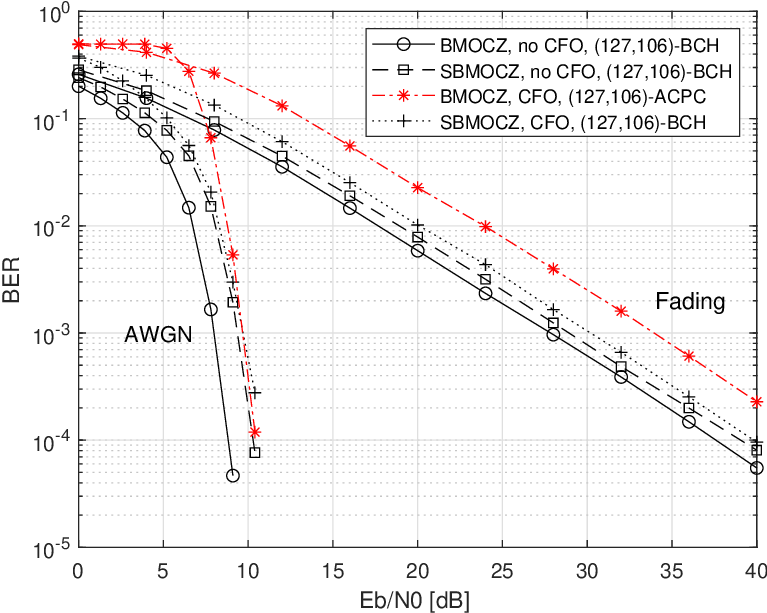}} \\
    	\subfloat[Block error rate curves. \label{subfig:coded_performance_bler}]{\includegraphics[width=3in]{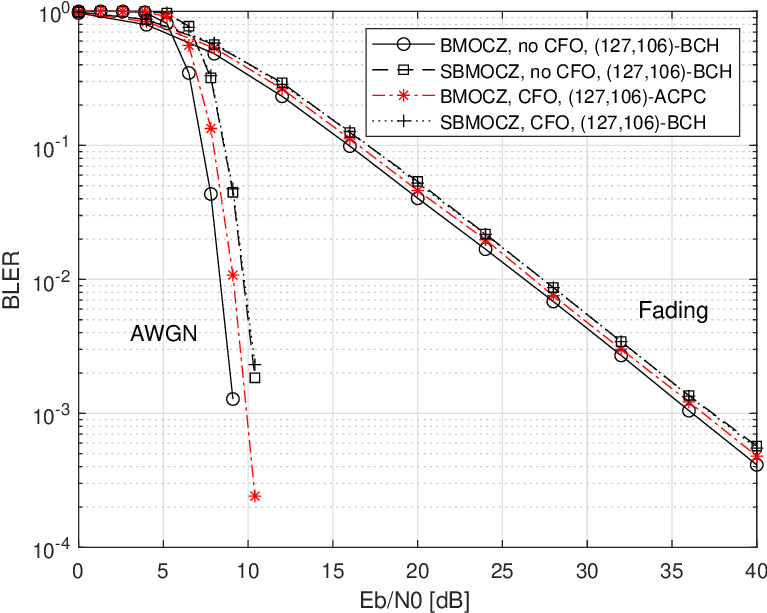}}
    	\caption{Comparison of coded schemes for $\numZeros=127$. The \ac{SBMOCZ} scheme is configured with $r_{\text{sb}}=1.0123$ and $\smooshFactor=0.0130$, while Huffman BMOCZ uses $r_{\text{hb}}=1.0123$.}
    	\label{fig:coded_performance}
    \end{figure}
   
\section{Numerical results} \label{sec:results}	
In this section, we compare the performance of \ac{SBMOCZ} to Huffman \ac{BMOCZ} in \ac{AWGN} and flat-fading channels. The noise coefficients $\noiseSeq\in\complexnumbers^{K+1}$ are drawn from $\complexnormal\left(0,\sigma_n^2\right)$, while the channel coefficient $h \in \complexnumbers$ is drawn from $\complexnormal\left(0,1\right)$ for the fading channel. To account for varying noise conditions, the noise variance $\sigma_n^2$ is computed for different $\ebno$. For simulations using a \ac{CFO}, we draw $\psi$ from the uniform distribution $\unif_{[0, 2\pi)}$ and apply the transformation in \eqref{eq:polynomial_cfo} to $Y(z)$. In each simulation, we randomly sample $\vecsym{m}\in\integers^K_2$ from the set of all $2^K$ possible binary messages. Furthermore, we utilize the \ac{dizet} decoder and set $N=1,024$ for the estimator in \eqref{eq:cfo_estimation_discrete}. The smooshing factor $\smooshFactor$ is selected using a parameter sweep, based on minimizing \ac{BER} under a \ac{CFO} in \ac{AWGN}.
	
   \subsection{Uncoded Error Rate Performance}
   To begin, we compare the performance of uncoded Huffman \ac{BMOCZ} to uncoded \ac{SBMOCZ} using $\smooshFactor = 0.0117$. For each simulation, we set $K=128$ and $\lambda=1/2$, and determine the radii using \eqref{eq:dizet_radius} for Huffman \ac{BMOCZ} and \eqref{eq:modified_radii} for \ac{SBMOCZ}. In Fig. \ref{fig:uncoded_performance}\subref{subfig:uncoded_performance_ber}, we plot the \ac{BER} curves for each scheme in \ac{AWGN} and fading channels. Without a \ac{CFO}, \ac{SBMOCZ} performs roughly $1.3$~dB worse than Huffman \ac{BMOCZ} in \ac{AWGN} and $0.85$~dB worse in fading. However, when a \ac{CFO} is introduced, Huffman \ac{BMOCZ} fails, as indicated by the red curve, while \ac{SBMOCZ} remains functional but loses $1.46$~dB in \ac{AWGN} and $2.92$~dB in fading relative to Huffman \ac{BMOCZ} without a \ac{CFO}. Observe that the \ac{BER} curve for \ac{SBMOCZ} with a \ac{CFO} starts at a higher point compared to the version without a \ac{CFO}, which occurs because the \ac{CFO} estimate fails at lower $\ebno$, resulting in a cascade of bit errors. In Fig. \ref{fig:uncoded_performance}\subref{subfig:uncoded_performance_bler}, we plot the \ac{BLER} for each scheme in \ac{AWGN} and fading channels. We observe that \ac{SBMOCZ} performs roughly the same with and without a \ac{CFO}, but is still approximately $1.5$~dB worse in \ac{AWGN} and $1$~dB in fading as compared to Huffman \ac{BMOCZ} without a \ac{CFO}. Again, we see that Huffman \ac{BMOCZ} without channel coding does not function under a \ac{CFO}.
    
   \subsection{Coded Error Rate Performance}
   For these simulations, we utilize a (127,106)-BCH code for \ac{SBMOCZ}. For Huffman \ac{BMOCZ}, we employ a (127,106)-BCH code absent a \ac{CFO} and a (127,106)-\ac{ACPC} under a \ac{CFO}. To ensure fair comparison, the \ac{ACPC} is implemented using a BCH code as the \ac{CPC}. We note that (127,106)-BCH corrects up to three bit errors, while (127,106)-\ac{ACPC} only corrects a maximum of two, sacrificing one bit of error correction to detect the cyclic shift $m\Delta{\phi}$. For (127,106)-\ac{ACPC}, we implement an \ac{IDFT}-based \ac{dizet} decoder with an oversampling factor of $Q=200$ to estimate and correct the fractional component $\psi_0$. Further details on the implementation of the \ac{ACPC} can be found in \cite{walk2020practical}. In each scheme, we set $K=127$ and $\lambda=1/2$, using $\smooshFactor = 0.0130$ for \ac{SBMOCZ}. Furthermore, we determine the radii using \eqref{eq:dizet_radius} for Huffman \ac{BMOCZ} and \eqref{eq:modified_radii} for \ac{SBMOCZ}.
    
   In Fig. \ref{fig:coded_performance}\subref{subfig:coded_performance_ber}, we plot the \ac{BER} for each scheme in \ac{AWGN} and fading channels. In \ac{AWGN}, \ac{SBMOCZ} and the \ac{ACPC} perform similarly at higher $\ebno$, while the \ac{ACPC} performs much worse at lower $\ebno$. For higher $\ebno$, both \ac{SBMOCZ} and the \ac{ACPC} perform roughly $1.6$~dB worse relative to coded Huffman \ac{BMOCZ} without a \ac{CFO}. However, in the fading channel, \ac{SBMOCZ} achieves a significant $4$~dB gain over the \ac{ACPC}. In Fig. \ref{fig:coded_performance}\subref{subfig:coded_performance_bler}, we plot the \ac{BLER} for each scheme in \ac{AWGN} and fading channels. We find that \ac{SBMOCZ} shows a loss of around $0.65$~dB in \ac{AWGN} and $0.6$~dB in fading compared to the \ac{ACPC}. Nevertheless, \ac{SBMOCZ} demonstrates a large \ac{BER} gain over the \ac{ACPC} in a fading channel.

\section{Concluding Remarks} \label{sec:conclusion}
In this study, we introduce a new smooshed \ac{BMOCZ} zero constellation called \ac{SBMOCZ} in which the angular separation between adjacent zeros, excluding the first and last, is reduced. By \textit{smooshing} the zeros closer together, we create a constellation gap that rotates under a \ac{CFO}, which enables the receiver to estimate and correct the rotation by identifying the gap's position. Compared to uncoded Huffman \ac{BMOCZ}, we find that uncoded \ac{SBMOCZ} functions under a \ac{CFO}, at the cost of a modest performance reduction without a \ac{CFO}. Against Huffman \ac{BMOCZ} with an \ac{ACPC}, we find that coded \ac{SBMOCZ} achieves a $4$~dB \ac{BER} gain in the fading channel, with comparable performance in other scenarios. Future work will focus on optimization of \ac{SBMOCZ} constellation parameters, such as the smooshing factor and the radii for each zero. Moreover, we will explore alternative decoding approaches for \ac{SBMOCZ}.

\appendices

\section{Radius Derivation for SBMOCZ} \label{app:sbmoczradii}
In \ac{SBMOCZ}, the radial seperation between a conjugate-reciprocal zero pair is given by
$d_{\rm cp} = r_{\text{sb}} - r_{\text{sb}}^{-1}$. The minimum separation between consecutive zeros, expressed as a function of $\smooshFactor$, is determined using the chord length formula, which yields $d_{\rm az}=2 r_{\text{sb}}^{-1} \sin\left(\frac{2\pi - \smooshFactor}{2K}\right)$.
To maximize the Euclidean distance between next-neighbor zero pairs, we equate $d_{\rm cp}$ and $d_{\rm az}$. However, since radial separation has a greater impact on \ac{BER} than angular separation \cite{walk2019principles}, we reintroduce the trade-off factor $\lambda \in \left(0,1\right]$ and obtain
\begin{equation} \label{eq:rad_der_3} \tag{1a}
	\begin{aligned}
		d_{\rm cp} &= \lambda d_{\rm az} \\
		r_{\text{sb}} - r_{\text{sb}}^{-1} &= 2 \lambda r_{\text{sb}}^{-1} \sin\left(\frac{2\pi - \smooshFactor}{2K}\right) \;.
	\end{aligned}
\end{equation} Solving for $r_{\text{sb}}$ gives
\begin{equation} \label{eq:rad_der_4}\tag{2a}
r_{\text{sb}} = \sqrt{1 + 2\lambda \sin\left(\frac{2\pi - \smooshFactor}{2K}\right)} \;.
\end{equation}

\section{Proof of Lemma 1} \label{app:lemma}
Let $\zeroVector=[r_{\text{sb}} \e^{\j\phi_0}, r_{\text{sb}} \e^{\j\phi_1},\dots,r_{\text{sb}} \e^{\j\phi_{K-1}}] \in \complexnumbers^\numZeros$ be the zeros of an \ac{SBMOCZ} polynomial $X(z)$. Evaluating on the unit circle and taking the squared magnitude yields
\begin{equation} \label{eq:lem1} \tag{1b}
	\begin{aligned}
		|X(\e^{-\j\theta})|^2 = |x_K|^2\prod_{k=0}^{K-1}\left|\e^{-\j\theta}-r_{\text{sb}}\e^{\j\phi_k}\right|^2 \\
		= |x_K|^2\prod_{k=0}^{K-1}f(\theta,\phi_k) \; .
	\end{aligned}
\end{equation} Using Euler's identities, we have
\begin{equation} \label{eq:lem2} \tag{2b}
	\begin{aligned}
		f(\theta,\phi_k) = \left|\e^{-\j\theta}-r_{\text{sb}}\e^{\j\phi_k}\right|^2= 1+r_{\text{sb}}^2-2r_{\text{sb}}\cos(\theta+\phi_k) \; .
	\end{aligned}
\end{equation}
Consider a case where $K$ is even. Since an \ac{SBMOCZ} constellation is symmetric about the real axis, we can rewrite this product as
\begin{equation} \label{eq:lem3} \tag{3b}
	\begin{aligned}
		|X(\e^{-\j\theta})|^2 = |x_K|^2 \prod_{k=0}^{\frac{K}{2}-1} f(\theta,\phi_k) \prod_{k=0}^{\frac{K}{2}-1} f(\theta,-\phi_k) \; .
	\end{aligned}
\end{equation} Observe that negating $\theta$ does not change each product term in \eqref{eq:lem3}, i.e., $f(\theta,\phi_k) f(\theta,-\phi_k) = f(-\theta,\phi_k) f(-\theta,-\phi_k)$. Therefore, we have 
\begin{equation} \label{eq:lem4} \tag{4b}
	\begin{aligned}
		\left|X(\e^{-\j\theta})\right|^2 = \left|X(\e^{\j\theta})\right|^2 \; .
	\end{aligned}
\end{equation} When $K$ is odd, the symmetry in \ac{SBMOCZ} still holds, with a single zero lying on the negative real axis. In this case, we can express \eqref{eq:lem1} as
\begin{equation} \label{eq:lem5} \tag{5b}
	\begin{aligned}
		|X(\e^{-\j\theta})|^2 = |x_K|^2 f(\theta,\pi) \prod_{k=0}^{\lfloor\frac{K}{2}\rfloor-1} f(\theta,\phi_k) \prod_{k=0}^{\lfloor\frac{K}{2}\rfloor-1} f(\theta,-\phi_k) \; .
	\end{aligned}
\end{equation} Observing that the term $\cos(\theta+\pi)=-\cos(\theta)$ in \eqref{eq:lem2} entails $f(\theta,\pi)=f(-\theta,\pi)$, we again arrive at \eqref{eq:lem4}. By Corollary~\ref{corr:dtftauto}, it follows that \eqref{eq:lem4} holds for any \ac{SBMOCZ} polynomial $X(z)$.
 
\bibliographystyle{IEEEtran}
\bibliography{references}

\end{document}

%% file: acronyms.tex
\acrodef{WSN}{wireless sensor network}
\acrodef{USRP}{universal software radio peripheral}
\acrodef{SN}{sensor node}
\acrodef{FC}{fusion center}
\acrodef{MAC}{multiple-access channel}
\acrodef{FL}{federated learning}
\acrodef{ED}{edge device}
\acrodef{CS}{compressed sensing}
\acrodef{ES}[BS]{base station}
\acrodef{DCN}{data center network}
\acrodef{RIS}{reconfigurable intelligent surfaces}
\acrodef{IMC}{in-memory computing}
\acrodef{FPGA}{field-programmable gate array}
\acrodef{SDR}{software-defined radio}
\acrodef{PS}{processing system}
\acrodef{SS}{soft synchronization}
\acrodef{IQ}{in-phase/quadrature}
\acrodef{IP}{intellectual property}
\acrodef{DMA}{direct-memory access}
\acrodef{RAM}{random access memory}
\acrodef{CC}{companion computer}
\acrodef{FEE}{function estimation error}
\acrodef{MSK}{minimum-shift keying}
\acrodef{TDMA}{time-domain multiple access}
\acrodef{PLNC}{physical-layer network coding}
\acrodef{UAV}{unmanned aerial vehicle}
\acrodef{LoRa}{Long-Range}
\acrodef{DC}{direct-current}
\acrodef{DAC}{digital-to-analog converter}
\acrodef{ADC}{anlog-to-digital converter}
\acrodef{CS}{complementary sequence}
\acrodef{GCP}{Golay complementary pair}
\acrodef{ANF}{algebraic normal form}
\acrodef{AACF}{aperiodic auto-correlation function}
\acrodef{AACFs}{aperiodic auto-correlation functions}
\acrodef{RM}{Reed-Muller}
\acrodef{MOCZ}{modulation on conjugate-reciprocal zeros}
\acrodef{BMOCZ}{binary modulation on conjugate-reciprocal zeros}
\acrodef{ACPC}{affine cyclically permutable code}
\acrodef{dizet}[DiZeT]{direct zero-testing}
\acrodef{DTFT}{discrete-time Fourier transform}

\acrodef{PUCCH}{physical uplink control channel}
\acrodef{PRACH}{physical random access channel}

\acrodef{OBO}{output-power back-off}
\acrodef{ACLR}{adjacent-channel-leakage ratio}

\acrodef{LDPC}{low-density parity check}

\acrodef{PDF}{probability density function}
\acrodef{CDF}{cumulative distribution function}
\acrodef{CCDF}{complementary cumulative distribution function}

\acrodef{TBMA}{type-based multiple access}

\acrodef{MSFE}{mean-squared function error}
\acrodef{FEE}{function-estimation error}
\acrodef{CER}{computation error rate}
\acrodef{BCER}{block-computation error rate}
\acrodef{CFO}{carrier frequency offset}
\acrodef{TO}{time offset}
\acrodef{PO}{phase offset}
\acrodef{RSSI}{received signal strength  information}

\acrodef{STLC}{space-time line code}
\acrodef{CCI}{co-channel interference}
\acrodef{CSIT}[CSIT]{\ac{CSI} at the transmitter}
\acrodef{CSIR}[CSIR]{\ac{CSI} at the receiver}
\acrodef{MIMO}{multiple-input multiple-output}
\acrodef{PC}{phase correction}
\acrodef{ZF}{zero-forcing}
\acrodef{ANOVA}{analysis of variance}

\acrodef{PCA}{principal component analysis}
\acrodef{TIG}{Technical Interest Group}

\acrodef{FSK}{frequency-shift keying}
\acrodef{PPM}{pulse-position modulation}
\acrodef{PAM}{pulse-amplitude modulation}

\acrodef{MRC}{maximum-ratio combining}
\acrodef{HP}{hard-coded participation}
\acrodef{HPA}{hard-coded participation with absentees}
\acrodef{SP}{soft-coded participation}
\acrodef{FSK-MV}{\ac{FSK}-based \ac{MV}}
\acrodef{RF}{radio-frequency}
\acrodef{MF}{matched filter}
\acrodef{PPM}{pulse-position modulation}
\acrodef{CSK}{chirp-shift keying}
\acrodef{PPM-MV}[PPM-MV]{\ac{PPM}-based \ac{MV}}
\acrodef{DFT-s-OFDM}{discrete Fourier transform-spread orthogonal frequency division multiplexing}
\acrodef{SC}{single-carrier}
\acrodef{SGD}{stochastic gradient descent}
\acrodef{signSGD}{sign stochastic gradient descent}

\acrodef{SL}{split learning}
\acrodef{SNR}{signal-to-noise ratio}
\acrodef{RMSE}{root-mean-squared error}
\acrodef{OFDM}{orthogonal frequency division multiplexing}
\acrodef{DFT}{discrete Fourier transform}
\acrodef{PSK}{phase-shift keying}
\acrodef{QAM}{quadrature amplitude modulation}
\acrodef{QPSK}{quadrature phase-shift keying}
\acrodef{PMEPR}{peak-to-mean envelope power ratio}
\acrodef{BER}{bit error rate}
\acrodef{SNR}{signal-to-noise ratio}
\acrodef{PSD}{power spectral density}
\acrodef{SE}{spectral efficiency}
\acrodef{CP}{cyclic prefix}
\acrodef{AWGN}{additive white Gaussian noise}
\acrodef{CFR}{channel frequency response}
\acrodef{CIR}{channel impulse response}
\acrodef{MMSE}{minimum mean-squared error}
\acrodef{LMMSE}{linear minimum mean-squared error}
\acrodef{BPSK}{binary phase shift keying}
\acrodef{BPSK}{quadrature phase shift keying}
\acrodef{BLER}{block error rate}
\acrodef{PHY}{physical layer}
\acrodef{PA}{power amplifier}
\acrodef{IDFT}{inverse discrete Fourier transform}
\acrodef{DoF}{degrees-of-freedom}
\acrodef{IoT}{Internet of Things}
\acrodef{mMTC}{massive machine-type communication}
\acrodef{URLLC}{ultra-reliable low-latency communication}
\acrodef{FDE}{frequency-domain equalization}
\acrodef{RF}{radio-frequency}
\acrodef{IM}{index modulation}
\acrodef{MF}{matched filter}
\acrodef{PPM}{pulse-position modulation}

\acrodef{MSE}{mean-squared error}
\acrodef{MRT}{maximum-ratio transmission}
\acrodef{ERC}{equal-ratio combining}
\acrodef{BAA}{broadband analog aggregation}
\acrodef{OBDA}{one-bit broadband digital aggregation}
\acrodef{FEEL}{federated edge learning}
\acrodef{FL}{federated learning}
\acrodef{UL}{uplink}
\acrodef{OAC}{over-the-air computation}
\acrodef{TCI}{truncated-channel inversion}
\acrodef{MV}{majority vote}
\acrodef{CNN}{convolution neural network}
\acrodef{ReLU}{rectified-linear unit}
\acrodef{CSI}{channel state information}
\acrodef{PAPR}{peak-to-average power ratio}
\acrodef{SC}{single-carrier}
\acrodef{iid}[IID]{independent and identically distributed}
\acrodef{RMS}{root-mean-square}
\acrodef{4G}{fourth generation}
\acrodef{5G}{Fifth Generation}
\acrodef{6G}{Sixth Generation}
\acrodef{NR}{New Radio}
\acrodef{LTE}{Long-Term Evolution}
\acrodef{OFDMA}{orthogonal frequency division multiple access}
\acrodef{HARQ}{hybrid automatic repeat request}
\acrodef{D2D}{Device-to-Device}
\acrodef{NOMA}{non-orthogonal multiple access}
\acrodef{OMA}{orthogonal multiple access}
\acrodef{IMT}{International Mobile Telecommunications}
\acrodef{ITU}{International Telecommunication Union}

\acrodef{PDP}{power-delay profile}
\acrodef{TBMA}{type-based multiple access}
\acrodef{ISI}{intersymbol interference}
\acrodef{MLSE}{maximum likelihood sequence estimator}
\acrodef{LTI}{linear time-invariant}
\acrodef{ISAC}{integrated sensing and communication}
\acrodef{ML}{machine learning}
\acrodef{DL}{deep learning}
\acrodef{AE}{autoencoder}
\acrodef{AEs}{autoencoders}
\acrodef{MLP}{multi-layer perceptron}

\acrodef{ICI}{inter-carrier interference}

\acrodef{SBMOCZ}{smooshed binary modulation on conjugate-reciprocal zeros}
\acrodef{UWB}{ultra-wideband}
\acrodef{CPC}{cyclically permutable code}
\acrodef{NSF}{National Science Foundation}

%% file: variables.tex
\newcommand{\herm}{^\text{H}}
\newcommand{\vecsym}[1]{\boldsymbol{\rm{#1}}}

\def\j{{\rm j}}
\def\e{{\rm e}}

\def\vector{\vecsym{v}}

\def\complexnormal{\mathcal{CN}}

\def\complexnumbers{\mathbb{C}}
\def\realnumbers{\mathbb{R}}
\def\integers{\mathbb{Z}}

\def\zeroCodebook{\mathscr{Z}}

\def\radius{R}
\def\numZeros{K}
\def\zeroIndex{k}
\def\zero{\alpha}

\def\polySeqTX{\vecsym{x}}
\def\polySeqRX{\vecsym{y}}

\def\zerosAndTaps{N_{\rm z}}

\def\zeroVector{\boldsymbol{\alpha}}

\def\filter{\vecsym{h}}
\def\numTaps{L}
\def\noiseSeq{\vecsym{w}}

\def\realcomponent[#1]{\text{Re}\left(#1\right)}
\def\imaginarycomponent[#1]{\text{Im}\left(#1\right)}

\def\complexity{\mathcal{O}}

\def\ebno{E_{\rm b}/N_0}

\def\messageSeq{\vecsym{m}}

\def\estMessageBit{\hat{m}}

\def\aacfindex{\ell}

\def\smooshFactor{\zeta}


%% file: main.bbl
\begin{thebibliography}{10}
\providecommand{\url}[1]{#1}
\csname url@samestyle\endcsname
\providecommand{\newblock}{\relax}
\providecommand{\bibinfo}[2]{#2}
\providecommand{\BIBentrySTDinterwordspacing}{\spaceskip=0pt\relax}
\providecommand{\BIBentryALTinterwordstretchfactor}{4}
\providecommand{\BIBentryALTinterwordspacing}{\spaceskip=\fontdimen2\font plus
\BIBentryALTinterwordstretchfactor\fontdimen3\font minus
  \fontdimen4\font\relax}
\providecommand{\BIBforeignlanguage}[2]{{%
\expandafter\ifx\csname l@#1\endcsname\relax
\typeout{** WARNING: IEEEtran.bst: No hyphenation pattern has been}%
\typeout{** loaded for the language `#1'. Using the pattern for}%
\typeout{** the default language instead.}%
\else
\language=\csname l@#1\endcsname
\fi
#2}}
\providecommand{\BIBdecl}{\relax}
\BIBdecl

\bibitem{nawaz2021backscatter}
S.~J. Nawaz, S.~K. Sharma, B.~Mansoor, M.~N. Patwary, and N.~M. Khan,
  ``Non-coherent and backscatter communications: Enabling ultra-massive
  connectivity in {6G} wireless networks,'' \emph{IEEE Access}, vol.~9, pp.
  38\,144--38\,186, 2021.

\bibitem{witrisal2009noncoherent}
K.~Witrisal, G.~Leus, G.~J. Janssen, M.~Pausini, F.~Troesch, T.~Zasowski, and
  J.~Romme, ``Noncoherent ultra-wideband systems,'' \emph{IEEE Signal
  Processing Magazine}, vol.~26, no.~4, pp. 48--66, 2009.

\bibitem{xu2019tradeoffs}
C.~Xu, N.~Ishikawa, R.~Rajashekar, S.~Sugiura, R.~G. Maunder, Z.~Wang, L.-L.
  Yang, and L.~Hanzo, ``Sixty years of coherent versus non-coherent tradeoffs
  and the road from {5G} to wireless futures,'' \emph{IEEE Access}, vol.~7, pp.
  178\,246--178\,299, 2019.

\bibitem{walk2019principles}
P.~Walk, P.~Jung, and B.~Hassibi, ``{MOCZ} for blind short-packet
  communication: Basic principles,'' \emph{IEEE Transactions on Wireless
  Communications}, vol.~18, no.~11, pp. 5080--5097, 2019.

\bibitem{huggins2024optimal}
P.~Huggins and A.~{\c{S}}ahin, ``On the optimal radius and subcarrier mapping
  for binary modulation on conjugate-reciprocal zeros,'' in \emph{Proc. IEEE
  Military Communications Conference (MILCOM)}, 2024, pp. 1--6.

\bibitem{walk2020practical}
P.~Walk, P.~Jung, B.~Hassibi, and H.~Jafarkhani, ``{MOCZ} for blind
  short-packet communication: Practical aspects,'' \emph{IEEE Transactions on
  Wireless Communications}, vol.~19, no.~10, pp. 6675--6692, 2020.

\bibitem{asahin2024majority}
A.~{\c{S}}ahin, ``Over-the-air majority vote computation with modulation on
  conjugate-reciprocal zeros,'' \emph{IEEE Transactions on Wireless
  Communications}, vol.~23, no.~11, pp. 17\,714--17\,726, 2024.

\bibitem{dehkordi2023integrated}
S.~K. Dehkordi, P.~Jung, P.~Walk, D.~Wieruch, K.~Heuermann, and G.~Caire,
  ``Integrated sensing and communication with {MOCZ} waveform,'' \emph{arXiv
  preprint arXiv:2307.01760}, 2023.

\bibitem{ackroyd1970design}
M.~H. Ackroyd, ``The design of {H}uffman sequences,'' \emph{IEEE Transactions
  on Aerospace and Electronic Systems}, no.~6, pp. 790--796, 1970.

\bibitem{walk2017short}
P.~Walk, P.~Jung, and B.~Hassibi, ``Short-message communication and {FIR}
  system identification using {H}uffman sequences,'' in \emph{Proc. IEEE
  International Symposium on Information Theory (ISIT)}, 2017, pp. 968--972.

\end{thebibliography}
